\documentclass[3p]{elsarticle}

\usepackage{amsmath,amssymb}
\usepackage{epsfig,graphicx}

\usepackage[T1]{fontenc}
\usepackage{aecompl}

\begin{document}
\title{Observational constraints on cosmological superstrings}
\author[rvt]{O.S. Sazhina\corref{cor1}\fnref{fn1}}
\ead{cosmologia@yandex.ru}
\author[focal]{A.I. Mukhaeva\fnref{fn2}}
\ead{mukhaeva.alfya@yandex.ru}
\cortext[cor1]{Corresponding author}
\address[rvt]{Sternberg State Astronomical Institute of Lomonosov Moscow State University}
\address[focal]{Dubna State University}
\fntext[fn1]{RF 119992 Moscow, Universitetskii prospect 13 (SAI MSU); tel.: +7 495 9395006; fax: +7 495 9328841
}
\fntext[fn2]{RF Dubna, Universitetskaya st. 19}

\begin{abstract}
From theoretical point of view and not being in contradiction with current observational data, the cosmic strings should have fundamentally different origin and are characterized by wide range of energies. The paper is devoted to search for possible cosmological observational tests on superstring theory, among them to identification of observational characteristics to distinguish between cosmological superstring of different types. In the brane-world scenario with an assumption of creation of cosmological superstrings it was obtained the lower limit on the superstring tension as function of deficit angle.
\end{abstract}


\date{}
\maketitle

\section{Introduction}

Cosmic strings are one-dimensional objects of the cosmological scales (see \cite {cs-1994} and the refs. therein). These exotic structures are predicted by theory, but has not been found yet \cite{planck}. Cosmic strings may be purely topological entities (endless or infinite, and closed loops), formed as a result of phase transitions in the vacuum stages of expansion and cooling of the early universe, and hybrid topological and field configurations (for example, a string with monopols on its end and conglomerates of such elements, so-called ''necklaces''). 

It is well-based the cosmological superstring theory as theory of extended objects with astrophysical characteristics which are the results of a wide variety of interactions of multidimensional spaces (namely the brane-world scenario, \cite{fs-2005}). Search for cosmological superstrings by astrophysical methods requires preliminary assessment of the characteristics of possible observed candidates on cosmic strings, namely, upper limits on cosmic string linear density. There are two main groups of methods in putting such the limits.
\begin{itemize}
\item String network simulations (see \cite{vil} and refs. therein) and search for cumulative restriction for string angular power spectrum in CMB data, \cite{planck};
\item Direct search for individual strings, \cite{br} (statement of Canny algorithm for future CMB experiments), \cite{ss-2014} (method based on Haar convolution, i.e. MHF method).
\end{itemize}

The first group of the methods gives the upper limit to estimate the fraction of energy in string with respect to the total energy of the Universe. 

According to the current observational data, \cite{pa}, there is no reason to assume that the strings have to manifest themselves in the form of network only. The negative results in the optical data to analyse the gravitational lens statistics indicate that (if the strings exist) they should be in small amount. Thus, it becomes important the  technique, which is able to recognize a single string and evaluate its possible features.

From the point of view of the cosmic string theory the unique parameter which determines all string properties is it linear density. The linear density has to completely define the conical geometry of string space-time (the string deficit angle).  But the main statement of cosmic string observational quest is that it is impossible to establish an unambiguous relationship between the linear density and deficit angle. This problem appears due to the large difference in the methods of the real observations. 

In observational search of cosmic string we used two independent methods.
\begin{itemize} 
\item Search for the special temperature jumps in CMB anisotropy data which indicate the string candidates, \cite{ss-2014}; 
\item Verification these string candidates by optical surveys looking for the chains of particular gravitational lensing events.
\end{itemize}

To determine the string linear density by the CMB anisotropy data it is necessary to identify a single string by the evaluation of the magnitude of the temperature jump, \cite{ss-2014}. On the other hand, to determine the string linear density by method of gravitational lensing it is necessary to identify lens candidates  (with its own probabilistic weight) and to measure the apparent angular distance between them. The string linear density estimated by the CMB anisotropy data, and the string linear density estimated by distances between lens components, are independent and different, being obtained by different methods. 

Finally, it is important to note that the linear density of cosmological superstrings in brane-world scenario is pure theoretical parameter and differs from the string linear density estimated by the CMB anisotropy data, and from the string linear density estimated by distances between lens components.

The upper limit on the linear density of cosmological superstring could be the same as for ordinary cosmic strings (pure topological or hybrid configurations). This limit has been obtained by CMB anisotropy, \cite{ss-2014}. The lower limit on the string linear density in brane-world scenario could be obtained from the gravitational lensing optical data. 

The paper is devoted to search of the lower observational limit on string linear density in brane-world model.

The paper is organized as follows. In the Section 2 the status of superstrings in modern cosmology is briefly described. In the Section 3 the warp factor is considered as a parameter which could be in principle estimated by the known observational string characteristic (string deficit angle). In the fourth (the main) Section it is discussed the connection between 4D cosmological superstrings and 10D superstrings from observational point of view. It is established the interval for values of linear density of the strings on the disscused model.

\section{Superstrings as possible cosmological objects}

Theoretical studies of cosmic strings become more popular in modern cosmology. Classical topological strings are of great interest for understanding the processes of phase transitions of the vacuum in the early universe, to study the relic dark energy, and to study the symmetry violation of fundamental interactions. Cosmic strings could serve as unique indicators of composition and topological properties of post-inflationary epochs.

The existence of cosmic strings of different types, properties and origins are not in contradiction with the modern observational data on CMB anisotropy. Hybrid cosmic string models become of particular interest because they are preferred both in terms of simulations \cite{kibble-2015}, and from the point of view of the observational data on the CMB anisotropy \cite {ss-2014}. Interest in the cosmic strings is manifested by the superstring physics as such objects may be the only observational evidence for superstring models \cite{polchinski}.

Firstly superstring have been considered as possible cosmological objects in connection with an understanding of the fact that their energy can be significantly lower than the Planck one (about $10^{19}$ GeV). Thus, the tension of the cosmological superstrings would be comparable with the observational limits. This approach is implemented in the representation of brane-world (in a multi-dimensional space-time when additional spatial dimensions have certain properties). In the 4D space-time fundamental superstring can not be stretched to the cosmological scale, they ''tear'' and will became a system of microscopic superstrings. In the brane-world scenario an extra dimension reduction provides an existance of superstrings on the cosmological scales. Let us consider the cosmological model of type II of superstrings, the existing 10D space-time (the existence of other types of cosmological superstrings is likely in contradiction with the observational data; see \cite {polchinski} and refs. therein). In string theory the superstrings are not the only size-localized objects. It is not forbidden a two-dimensional membrane or their analogs of higher dimensions, called p-brane (p is the brane dimension); so the particle is a 0-brane, the string is 1-brane, etc. In the brane-world model it is introduced the concept of Dp-branes, which means that the p-brane satisfies the initial conditions of Dirichlet-type to fix  the ends (the same way as it is done in the solution of differential equations in partial derivatives). Thus in the multi-dimensional space-time there are fundamental closed strings (loops), and fundamental strings which are ended in the Dp-branes. The model also allows anti-Dp-branes: a brane and an anti-brane have equal and opposite charges, so they are attracted to each other. More generally, M-theory, too, contains the brane-world scenario. The common property of these models is that ordinary matter is concentrated on the hypersurface, the brane immersed in a space-time of higher dimension (called the bulk). Our universe may be one of such a brane. Gravity can be spread in the bulk (for review, see \cite{rubakov}).

Thus the superstring can naturally appear in brane-world scenario with energies that are comparable with observational constraints on cosmic string energies.

\section{Warp-factor in the 10D space-time}

One of the main motivations of brane-world scenario was to explain the large energy gap between Planck scale of gravity $10^{19}$ GeV and electroweak interactions, $10^{2}$ GeV. The model bases on the warp factor of the space-time. As in the general relativity the interval is defined
$$
ds^2 = dt^2 - dx^2,
$$
in the warp space-time it is introduced the modified interval:
$$
ds^2 = e^{-A(y)} \biggl( dt^2 - dx^2 \biggr) - dy^2.
$$
The $x$ is a three-dimensional spatial vector, the $y$ is coordinate (or coordinates) of additional compact dimension, $e^{- A(y)}$ is a positive function of the extra dimension, called warp factor (the function $A(y)>0$) . The physical meaning of this function is that it is the gravitational redshift in a compact extra dimension. In the simplest case of a 5D space-time the warping is introduced to define a hierarchy of energy scales. Thus, gravity, which is distributed in both the bulk and on the brane can have the Planck energy, while ordinary physics, whose particles and interactions are localized only on the brane, can be characterized by a much lower energies.

The idea of space-time warping extends to a greater number of additional compact dimensions (6, in the case of a 10D space-time). In the simplest models, the extra dimensions are spheres or tori, characterized by $A(y)=const$. There are more complicated models where warp factor $e^{-A(y)}$ strongly depends on the coordinate $y$. Thus, in the so-called throat of multi-dimensional space-time, this function is small, and far away from the throat is close to unity. For an observer located in the 4D space-time it means that the observed tension (or linear density) of the cosmological fundamental strings is
$$
\mu_0 = e^{-A(y)} \cdot \mu,
$$
where $\mu$ is defined in 10D space-time. In other words, $\mu_0$ comes from real observational data (from the CMB $\delta T/T$ upper limits or from the statistically averaged optical data $\Delta \Theta$), and $\mu$ is theoretical characteristic of the superstring. 

Therefore, the observed energy of the fundamental string could be much smaller than its energy in 10D space-time. 

It is important to note that in the brane-world scenario together with the birth of cosmological superstrings it could naturally be realized the inflation stage of the early universe (see \cite{polchinski}). Inflation occurs when the energy reaches into the so-called throat, i.e., at a low value function $e^{-A(y)}$. Inflation is carried out in the process of annihilation of a brane and an anti-brane. In this process branes and anti-branes of higher dimensions can produce branes of lower dimensions, including the fundamental strings. The string energy scale is not Planck one, it is much lower because the strings are created in areas of high warping. Thus, the inflation energy is directly linked with the energy of fundamental superstrings (being in frames of the brane-world scenario).

\begin{figure}[pH]
\centering\includegraphics[width=15.0cm, angle=90.0]{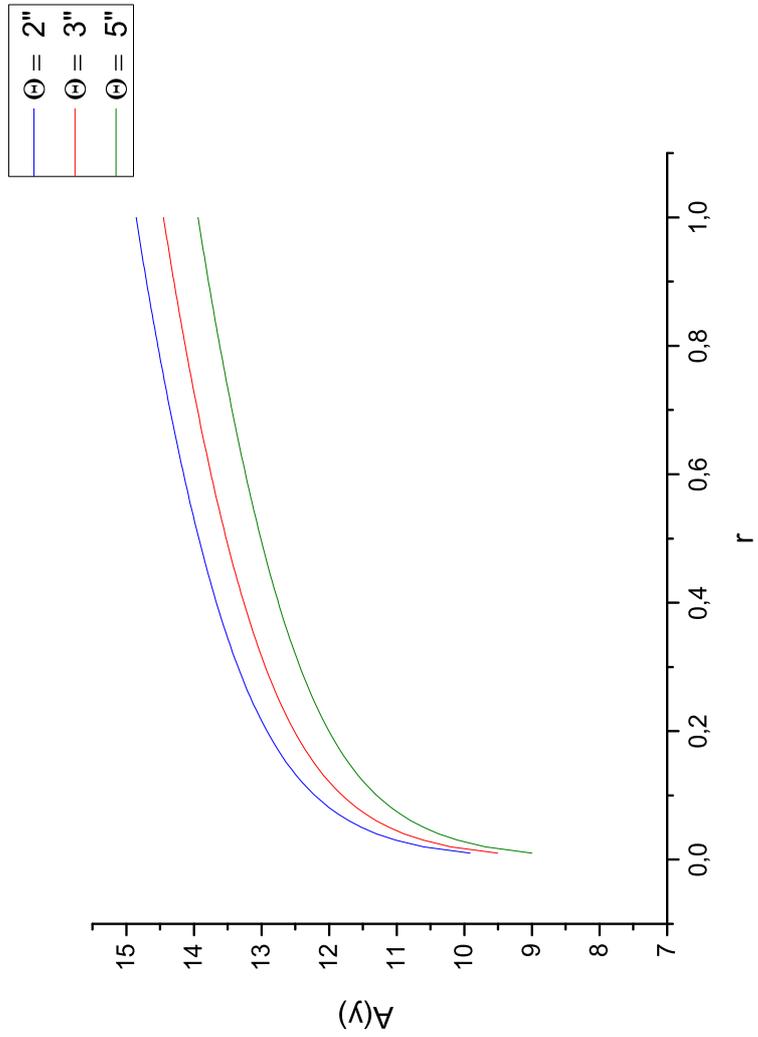}
\caption{\textit{The 3D diagram of the dependence of the warp factor function $A(y)$ on the deficit angle $\Delta \Theta$, measured in arcsec, and on the ratio of the Planck length to the length of the fundamental string $r = l_p/l_s$ is presented. The different colors correspond to different values of $ p, q$, describing the mixed state configuration of the string, consisting of $p$ $F$-strings and $q$ $D$-strings. For inflation models on the brane with the warp factor of the order of $10^{- 4}$, \cite{polchinski}, the energetically preferred configuration are those with the least number of the $F$- and $D$-strings.}}
\label{f1}
\end{figure}

\section{Connection between 4D cosmological superstrings and 10D superstrings}

The tension of the fundamental strings, observed in the usual 4D space-time, is related to its 10D tension proportionally to the warp factor:
\begin{eqnarray}
\mu_0 = e^{-A(y)} \cdot \mu_{p,q}^{(10)}.
\label{main}
\end{eqnarray}
Indices $p$ and $q$ represent the number of F- and D-strings respectively. As F-, and D-strings can be generated during inflation in the annihilation process of branes and anti-branes. F- and D-strings can combine to form one-dimensional cosmological objects called FD-strings (as visual analogy for the case, for example, $p = 1, q = 1$ there could be two lines approaching each other at a certain angle, and then twisted with one another into a single thread).

The cosmic string tension from the point of view of 4D observer depends on the string velocity and on the temperature amplitude of the CMB, \cite{ks}, \cite{s}:
\begin{eqnarray}
\frac{\delta T}{T} = \frac{8\pi G \mu_0}{c^2} \cdot \frac{\beta}{\sqrt{1-\beta^2}}.
\label{l}
\end{eqnarray}
Here $\beta$ is the component of cosmic string velocity, normalized to the speed of light, perpendicular to the line of sight, $ T = 2.73 $ is the temperature of the CMB, $\delta T = 40 \mu K$ is the observational limit on the maximal value of the CMB anisotropy generated by a single cosmic string, obtained by MHF method \cite{ss-2014}. 

For example in the simplest case of Nambu-Goto network simulations the string velocity is fixed: $\beta = 1/\sqrt{2}$  (\cite{ss-2014}, \cite{cs2010}). It is important to note that now the ability to search individual strings in the CMB anisotropy data provides the ability to find connection between the velocity $\beta$ and linear density $\mu_0$ of a string as a function of the upper limit on the anisotropy generated by the string:

\begin{eqnarray}
F(\beta, \mu_0) = \frac{\delta T}{T}. 
\label{fu}
\end{eqnarray}

On the other hand, one can use observational data of a completely different nature, from optical surves. Rewriting the expression for linear density in the Planck unit system ($\hbar = c = 1$):
\begin{eqnarray}
\mu_0 = \Delta \Theta \frac{1}{8\pi} \cdot \frac{1}{l_p^2},
\label{_mu}
\end{eqnarray}
where $l_p = \sqrt{G}$ is the Planck length, and $\Delta \Theta = 8\pi G \mu_0 / c^2$ is the exact theoretical definition of the string deficit angle. The deficit angle is the angle of the 3D space of the cone, which replaces the 3D Euclidean space of our universe in the presence of a cosmic string, \cite{vil}. From the point of view of an observer the deficit angle is proportional (in the case of a large cosmic string distance from the observer, asymptotically equal) to the angular distance between the components of pairs of gravitationally lensed images of distant galaxies lensed by a cosmic string. Therefore, in the approach of the model of remote string $\Delta\Theta$ is observational parameter, which should be estimated statistically analizing the chain of gravitational lenses along the path of the string.

Obviously, at different velocities $\beta$ with the same magnitude of the anisotropy the deficit angle should be different. Using both the radio and optical data, similarly as it was presented in (\ref{fu}), one can reconstruct the more probably string velocity $<\beta>$:
$$
<\beta> = \tilde{F} \biggl(\Delta\Theta, \frac{\delta T}{T} \biggr), 
$$
where in general $\Delta \Theta$ should be a statistical value coming from observational data,
$$
\Delta\Theta = <\Delta \Theta> \pm \frac{\sigma_{lens}}{\sqrt{N_{lens}}}. 
$$
Here $N_{lens}$ is the number of gravitational lenses along the proposed (by CMB data) position of a string and $\sigma_{lens}$ is the standard deviation of the measurement of the distances between the pairs. It will be done in future work (the analysis of gravitational lenses along the cosmig string candidate with $\delta T/ T = 40 \mu K$, \cite{ss-2014}, now in proggress).

A theoretical (nonstatistical) expectations of the dependence of $\beta$ and $\Delta \Theta$ using the simulations of the evolution of a network of cosmic strings allow the value of $\beta$ of 0.58 (see. Table \ref{_tab1}).

\begin{table} 
\begin{center}
\begin{tabular}{|c|c|c|c|c|c|c|} 
\hline
Properties/references        &      & \cite{b}       & \cite{a}       & \cite{b}  & \cite{a}   &   \\
                             &      & radiation ep.  & radiation ep.  & matter ep.& matter ep. &  \\
                             & & & & & & \\
\hline 
& & & & & & \\
$\beta$                      & 0.99 & 0.66           & 0.62           & 0.61      & 0.58       & 0.50  \\
& & & & & & \\
\hline
& & & & & & \\
$\Delta\Theta ('') $         & 0.43 & 3.47           & 3.86           & 3.97      & 4.29       & 5.29 \\
& & & & & & \\
\hline
\end{tabular}\caption{There are presented the values $\beta$, obtained by simulation of the dynamics of string networks (there are listed the authors of the respective papers). From the values of $\beta$ is calculated the deficit angle (in arcsec), which corresponds to the anisotropy of $40 \mu K $ (see text). The first and last columns of pairs $\beta$ and $\Delta \Theta$ are presented as limiting values that do not contradict the observational data. }\label{_tab1}
\end{center}
\end{table}

\begin{figure}[pH]
\centering\includegraphics[width=15.0cm, angle=90.0]{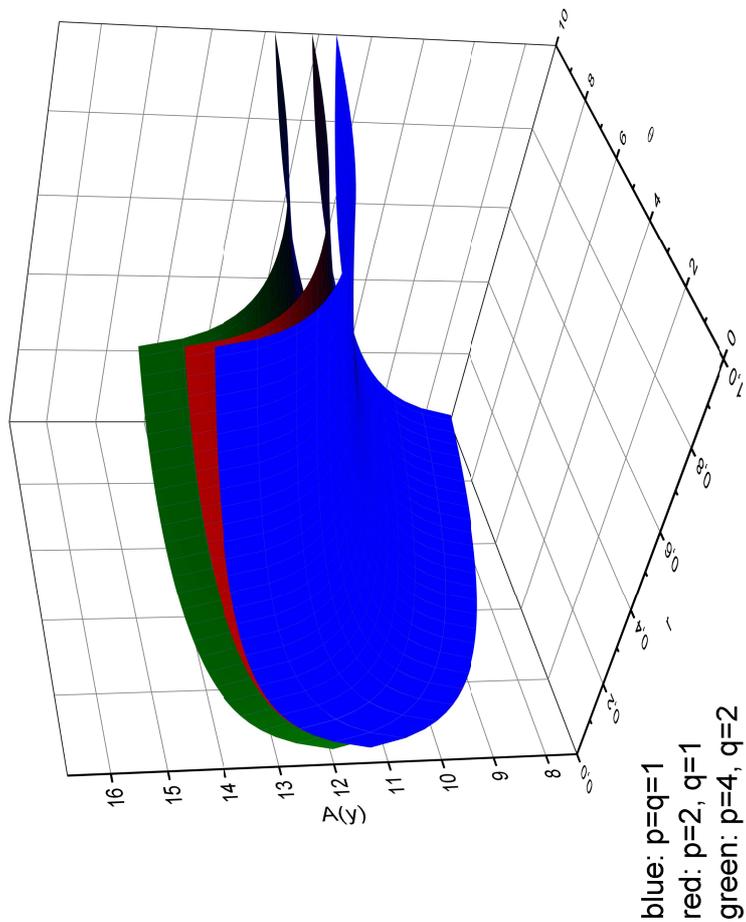}
\caption{\textit{The dependence of the warp factor function $A(y)$ of the additional dimension from the ratio of the Planck length to the length of the fundamental string $r = l_p/l_s$ (cross section at different $\Delta \Theta$ of the surfaces in Fig. (\ref{f1}) for $p=q=1$). When the length of the string is much larger than the Planck (model of large extra dimensions), fundamental strings are created during the inflation stage, which corresponds to the warp factor of the order of $10^{-4}$ or $A(y) \approx 9$. The values of the warp factor are significantly sensitive to observational parameter of a string -- deficit angle.}}
\label{f2}
\end{figure}

Now let us turn to (\ref{main}). The 10D tension is expressed by the formula:
\begin{eqnarray}
\mu_{p,q}^{(10)} = \frac{1}{2\pi l_s^2} \cdot \sqrt{p^2 + \frac{q^2}{g_s^2}},
\label{r}
\end{eqnarray}
where $g_s$ is the string coupling constant, $l_s$ is the lenght of the fundamental string (defined by its creation energy). 

In the 10D string theory, the coupling constant $g_s$ can be expressed in terms of the length of the fundamental string and Planck length and also through the 4D Newton gravitational constant:

$$
G = g_s^2 \alpha', 
$$
$$
l_s = \sqrt{\alpha'},
$$
$$
g_s^2 = \frac{G}{\alpha'} = \frac{G}{l_s^2} = \frac{l_p^2}{l_s^2},
$$
where $\alpha'$ is fundamental string parameter, which is called inclination angle, $[E^{-1}]$.

Let us suppose that from optical observation we obtained the value $<\Delta \Theta>$ and calculated $\sigma_{lens}$ as we discussed recently. Then, substituting in (\ref{main}) expressions (\ref{r}) and (\ref{_mu}): 
\begin{eqnarray}
\Delta\Theta \cdot \frac{1}{8\pi} \cdot \frac{1}{l_p^2} = e^{-A(y)} \cdot \frac{1}{2\pi} \cdot \frac{1}{l_s^2} \cdot \sqrt{p^2 + q^2 \frac{l_s^2}{l_p^2}}.
\label{_m}
\end{eqnarray}
On the left part of the equation (\ref{_m}) there is the observed deficit angle (by the optical method of the search of gravitational lensing events). On the right part of the equation (\ref{_m}) there is a function which depends on the theoretical linear density of FD-string. The theoretical linear density of the FD-string is limited by real observable characteristic (deficit angle). This gives us the ability to link the theoretical energy of FD-string creation with the real observed characteristics.

Transforming the last expression, it is obtained:
$$
e^{A(y)} = \frac{4}{\Delta \Theta} \cdot \biggl(\frac{l_p}{l_s}\biggr)^2 \cdot \sqrt{p^2 + q^2 \biggl(\frac{l_s}{l_p}\biggr)^2 }.
$$

The ratio of fundamental string length to the Planck length it is denote $r$: $l_p/l_s = r$. Then
$$
A(y) = \ln \biggl( \frac{4r^2 \sqrt{p^2+q^2/r^2}}{\Delta \Theta} \biggr).
$$
The function $A(y) > 0$, therefore, $r > r_{crit}$:
$$
r_{crit} = \frac{1}{\sqrt{2}p} \cdot \sqrt{ \sqrt{q^4+p^2 \biggl( \frac{\Delta\Theta}{2} \biggr)^2 } - q^2 }.
$$
Since the value  $(p \cdot \Delta\Theta/2) << 1$, up to the third order low
$$
r_{crit} = \frac{\Delta \Theta}{4q}.
$$

Therefore,
$$
\frac{l_p}{l_s} > \frac{\Delta \Theta}{4q} 
$$
or
$$
\frac{m_s}{m_p} > \frac{\Delta \Theta}{4q}.
$$
The relation between the string mass and the superstring theoretical tension (linear density) is done by ($c = \hbar=1$):
$$
\biggl( \frac{m_s}{m_p} \biggr)^2 = G\mu_{s}.
$$
Finally, we obtain a limit on the minimum possible tension of FD-string as follows:
$$
\biggl(\frac{\Delta\Theta}{4q}\biggr)^2 = \biggl(\frac{\Delta\Theta}{2''}\biggr)^2 \cdot \frac{10^{-10}}{16q^2} < G\mu_{s}.
$$
Note that the resulting restriction does not depend on the number of F-strings. The expression is consistent with those of other authors ($10^{-12}<G \mu_{s} <10^{-6}$, \cite{polchinski}) and refines them according to the observational data. For large, but acceptable by observations, deficit angle values (which corresponds to less string velocities) minimum value of the tension reaches a value of about $10^{-9}$.

For strings, moving at about light velocity, the minimum value of tension is about $10^{-12}$. Note also that all previously obtained estimations of cosmic string linear density were based on the cumulative assessment of the possible contribution of the spectrum of cosmic strings in the spectrum of the CMB anisotropy, while our estimates are based on an analysis of individual candidates for cosmic strings.

The minimal possible energy of FD-strings' formation (in GeV) is:
$$
\frac{\Delta \Theta}{4q} \cdot 1.2 \cdot 10^{19} = \biggr( \frac{\Delta \Theta}{2''} \biggr) \cdot \frac{3 \cdot 10^{13}}{q} < m_s.
$$.

Therefore the main characteristic of cosmological superstring is not a free parameter, as it was assumed until recently, \cite{polchinski}. Note also the significant difference between the lower limit of the transition from the $q = 1 $ to all $q> 1$. Thus, from the observational point of view can be distinguished the situation of having only one D-string, or many. The lower limit is still small: so for $\beta = 0.9$ the lower limit of $G \mu_{s} = 2.9 \cdot 10^{-13}$, and for $\beta = 0.5$ the lower limit of $G \mu_{s} = 4.4 \cdot 10^{- 11} $. With the growth of q the lower limit obviously decreases. The range for the value of the tension of the resulting combinations of FD-string is insensitive to the number of F-strings.

As it was mentioned above (see. Table \ref{_tab1}), for a given value of $ \delta T / T$ (which has been defined by analysis of CMB anisotropy data, \cite{ss-2014}) there is a unique relationship between the string velocity $\beta$ and deficit angle $ \Delta \Theta $. For classical strings (not superstrings) definition by independent methods (for example, by searching for strings of gravitational lens pairs of distant galaxies) of the string deficit angle allows uniquely restored its velocity, and its tension $\mu$. In the case of superstrings by knowing the deficit angle, one could restore the warp factor $e^{-A(y)} $ and determine the creation energy of the strings.

\section{Conclusions}
If the brane-world scenario is correct, then the annihilation of the brane and antibrane creates FD-strings. Using the 4D observational features of the possible string candidates one should put a lower limit on the superstring linear density $\mu_{s}$: 
$$
\biggl(\frac{\Delta\Theta}{2''}\biggr)^2 \cdot \frac{10^{-10}}{16q^2} < G\mu_{s},
$$
where $q$ is the number of D-strings and $\Delta\Theta$ is the string deficit angle (in arcsec). 

Through analysis of this string candidate in the optical surveys there were found lens candidates which determine the average value of string deficit angle $\Delta \Theta$. Then, what is the main conclusion of the paper, it is possible to calculate a lower limit on the string linear density (under the assumption of the production of the FD-strings). This limit is derived from the simple condition of positiveness of the warp factor.

The theory of the cosmological string becomes very rich, they could have different origin and are characterized by wide range of energies. 

Due to new projects on gravitational lensing studies and the CMB data analysis some energy bands for cosmic strings to exist are closing. So, there is less hope to observe so-called light strings (with energies less than $10^{15}$ -- $10^{16}$ GeV) by CMB data and also by method of gravitational lensing. However, some new possibilities for strings are opening. The cosmological superstring could be characterized by astrophysical parameter (the deficit angle of the string). 

The fixed value of CMB anisotropy induced by a string and known deficit angle (by gravitation lensing data) could uniquelly restore the string velocity and string linear density. In the case of cosmological superstring one could restore the warp factor and determine the string creation energy.

\section*{Acknowledgments}

The authors thank Prof. M.V. Sazhin for useful discussions.

\end{document}